\documentclass{amsproc}

\theoremstyle{definition}

\theoremstyle{remark}

\numberwithin{equation}{section}
                                                                                
\begin{document}

\hfill hep-th/0502064

\vspace{0.2in}                                                                                
\title{Notes on Correlation Functions in (0,2) Theories*\footnote{*To appear in the Proceedings of the Conference on String Geometry, June 5-11, 2004, Snowbird, Utah.}}
                                                                                
\author{Eric Sharpe}
\address{Physics Department, University of Utah, 115 S. 1400 E. Suite 201,
Salt Lake City, Utah 84112-0830}
\curraddr{Physics Department,
University of Utah, 115 S. 1400 E. Suite 201,
Salt Lake City, Utah 84112-0830}
\email{ersharpe@math.utah.edu}
\thanks{}
                                                                                
%\author{Author Two}
%\address{Mathematical Research Section, School of Mathematical Sciences,
%Australian National University, Canberra ACT 2601, Australia}
                                                                                
%    General info
\subjclass{Primary 14J32, 14N35; Secondary 14J60}
\date{June 7, 2004.}
                                                                                
                                                                                
\keywords{Mirror symmetry, algebraic geometry}
                                                                                
\begin{abstract}
In this talk we shall review some recent work on generalizing
rational curve counting to perturbative heterotic theories.
\end{abstract}
                                                                                
\maketitle

\section{Introduction}

In this short note we shall review some recent work \cite{ks}
on certain correlation function calculations in perturbative
heterotic strings, generalizing the rational curve counting of
the A model to analogues of the $\overline{ {\bf 27} }^3$ coupling.
 
Part of the motivation for this work comes from attempts to
understand (0,2) mirror symmetry.  Recall that ordinary mirror
symmetry relates pairs of Calabi-Yau manifolds.
By contrast, (0,2) mirror symmetry relates Calabi-Yau manifolds
with holomorphic vector bundles:
\begin{displaymath}
\left( X_1, \mathcal{E}_1 \right) \: \Longleftrightarrow \:
\left( X_2, \mathcal{E}_2 \right)
\end{displaymath}
The Calabi-Yau manifolds $X_1$, $X_2$ need not be mirror in the
ordinary sense, except in the special case that $\mathcal{E}_i = TX_i$,
in which case (0,2) mirror symmetry reduces to ordinary mirror symmetry.

At present, (0,2) mirror symmetry is very poorly understood.

Recently, \cite{adamsbasusethi} studied (0,2) mirrors.
They applied ideas of \cite{horivafa,daveronen2} to (0,2) gauged
linear sigma models to make some predictions for (0,2) mirrors in some
relatively simple cases.  In particular, they made some physical predictions
for the results of heterotic rational curve counting.  One of the
motivations of \cite{ks} was to develop the technology required
to check those predictions directly.

In the first part of this talk we shall outline formally how the
computation of the relevant correlation functions can be
translated into mathematics.  To perform the indicated mathematical
computations will require not only compactifying certain moduli spaces
of rational curves, but also extending certain sheaves over those
compactifications.  In the second part of this talk we will outline
how gauged linear sigma models naturally provide mechanisms for doing
both.

The reader should note that the old vanishing results of
\cite{candelasetal,beasley,evaed} apply to spacetime superpotential
contributions involving gauge singlets, whereas by contrast,
the present note is devoted to spacetime superpotential contributions
involving only charged states.  We are interested in analogues of the
$\overline{ {\bf 27} }^3$ coupling, which receives rational curve
corrections, whereas \cite{candelasetal,beasley,evaed} study
${\bf 1}^3$ and ${\bf 27} - \overline{ {\bf 27} } - {\bf 1}$ couplings.

\section{Nearly topological field theories}

Recall that on the $(2,2)$ locus, the A model topological field
theory is a twist of the nonlinear sigma model which is amenable
to rational curve counting.  Specifically, the A model is defined
by twisting worldsheet fermions into worldsheet scalars and vectors
as follows \cite{edtft}:
\begin{displaymath}
\begin{array}{ccc}
\psi_+^i \: \in \: \Gamma_{ C^{\infty} }\left( \phi^* T^{1,0}X \right), & \: \:
\: &
\psi_-^i \: \in \: \Gamma_{ C^{\infty} }\left(
\overline{K}_{\Sigma} \otimes \left( \phi^* T^{0,1} X \right)^{\vee} \right), \\
\psi_+^{\overline{\imath}} \: \in \:
\Gamma_{ C^{\infty} }\left( K_{\Sigma} \otimes \left(
\phi^* T^{1,0} X \right)^{\vee} \right), & \: \: \: &
\psi_-^{\overline{\imath}} \: \in \: \Gamma_{ C^{\infty} }\left(
\phi^* T^{0,1} X \right).
\end{array}
\end{displaymath} 
The heterotic analogue of the A model is a twist of the $(0,2)$ nonlinear
sigma model in which the fermions couple to bundles as follows:
\begin{displaymath}
\begin{array}{ccc}
\psi_+^i \: \in \: \Gamma_{ C^{\infty} }\left( \phi^* T^{1,0} X \right), &
\: \: \: & \lambda_-^a \: \in \: \Gamma_{ C^{\infty} }\left(
\overline{K}_{\Sigma} \otimes \left( \overline{ \mathcal{E} } \right)^{\vee} 
\right), \\
\psi_+^{\overline{\imath}} \: \in \: \Gamma_{ C^{\infty} } \left(
K_{\Sigma} \otimes \left( \phi^* T^{1,0}X \right)^{\vee} \right), 
& \: \: \: &
\lambda_-^{\overline{a}} \: \in \: \Gamma_{ C^{\infty} }\left(
\overline{ \mathcal{E} } \right),
\end{array}
\end{displaymath}
where $\mathcal{E}$ is a holomorphic vector bundle on $X$, which
will be assumed to obey the anomaly cancellation condition
$\mbox{ch}_2(\mathcal{E}) = \mbox{ch}_2(TX)$.
Although both left- and right-movers have been twisted,
the theory defined by the twisting above is not a topological
field theory, since the worldsheet does not have supersymmetry
on left-movers.

The states of the A model are counted by $H^{p,q}(X)$,
and they obey the symmetry
\begin{displaymath}
H^{p,q}(X) \: \cong \: H^{n-p,n-q}(X)^*
\end{displaymath}
for $X$ compact and of dimension $n$.

The RR states of the (0,2) theory are counted by 
sheaf cohomology $H^q(X, \Lambda^p \mathcal{E}^{\vee})$, and they
obey
\begin{displaymath}
H^q\left(X, \Lambda^p \mathcal{E}^{\vee} \right) \: \cong \:
H^{n-q}\left(X, \left( \Lambda^{r-p} \mathcal{E}^{\vee} \right)
\otimes \left( \Lambda^{top} \mathcal{E} \otimes K_X \right) \right)^*
\end{displaymath}
for $X$ compact of dimension $n$, and $\mathcal{E}$ of rank $r$.

We will make the assumption that $\Lambda^{top} \mathcal{E}^{\vee} 
\otimes K_X$ throughout this lecture, in addition to the usual
anomaly cancellation condition $\mbox{ch}_2(\mathcal{E}) =
\mbox{ch}_2(TX)$.  One reason for this condition is that it reproduces
the same symmetry property of states in the (0,2) theory that is possessed
by the A model.  We shall see that it also plays a critical role
in computing correlation functions.

Also note that on the (2,2) locus, where $\mathcal{E} = TX$,
$\Lambda^{top} \mathcal{E}^{\vee} = K_X$ automatically.

\section{Classical correlation functions}

Before discussing the (0,2) model, let us first review classical
correlation functions in the A model.

In the A model, for $X$ compact and $n$-dimensional,
we have $n$ $\chi^i$ zero modes, $n$ $\chi^{\overline{\imath}}$
zero modes, and the space of bosonic zero modes is $X$ itself.
Thus,
\begin{displaymath}
< \mathcal{O}_1 \cdots \mathcal{O}_m > \: \sim \:
\int_X H^{p_1,q_1}(X) \wedge \cdots \wedge H^{p_m,q_M}(X).
\end{displaymath}
(Our notation is concise but sloppy -- we wish to indicate an 
integral of a product of elements of cohomology classes,
so we use the cohomology group itself to denote an element.)

The selection rules from the left- and right-moving $U(1)$ anomalies
give the constraint
\begin{displaymath}
\sum p_i \: = \: \sum q_i \: = \: n
\end{displaymath}
thus, the correlation function above can be expressed in the form
\begin{displaymath}
< \mathcal{O}_1 \cdots \mathcal{O}_m > \: \sim \:
\int_X \left( \mbox{top form} \right).
\end{displaymath}

We will see the same pattern throughout -- correlation functions
will always be expressable as integrals of top-forms.

In the (0,2) model, classically we have $n$ $\psi_+^{\overline{\imath}}$
zero modes and $r$ $\lambda^a$ zero modes, where $r$ is the rank of
$\mathcal{E}$.
Thus, we can write
\begin{displaymath}
< \mathcal{O}_1 \cdots \mathcal{O}_m > \: \sim \:
\int_X H^{q_1}\left( X, \Lambda^{p_1} \mathcal{E}^{\vee} \right) \wedge
\cdots \wedge
H^{q_m}\left(X, \Lambda^{p_m} \mathcal{E}^{\vee} \right).
\end{displaymath}
The selection rules from the left- and right-moving $U(1)$ anomalies
give the constraints
\begin{displaymath}
\sum q_i \: = \: n, \: \: \:
\sum p_i \: = \: r
\end{displaymath}
thus,
\begin{displaymath}
< \mathcal{O}_1 \cdots \mathcal{O}_m > \: \sim \:
\int_X H^{top}\left(X, \Lambda^{top} \mathcal{E}^{\vee} \right).
\end{displaymath}
Now, elements of $H^{top}(X, \Lambda^{top} \mathcal{E}^{\vee})$
are not top-forms in general.
However, recall we have the constraint $\Lambda^{top} \mathcal{E}^{\vee}
\cong K_X$, and elements of $H^{top}(X, K_X)$ {\it are} top forms.
Thus, so long as $\Lambda^{top} \mathcal{E}^{\vee} \cong K_X$,
we can write
\begin{displaymath}
< \mathcal{O}_1 \cdots \mathcal{O}_m > \: \sim \:
\int_X \left( \mbox{top form} \right)
\end{displaymath}
classically in the (0,2) model.

\section{Worldsheet instantons}

First, let us review the A model.
The A model TFT localizes on holomorphic maps, so the moduli space
of bosonic zero modes is a moduli space of worldsheet instantons
$\mathcal{M}$.  We shall assume that $\mathcal{M}$ is smooth,
and will review relevant compactifications later.

In the (0,2) model, the bundle $\mathcal{E}$ on $X$ induces a sheaf
$\mathcal{F}$ on $\mathcal{M}$, given by $\mathcal{F} \equiv \pi_* \alpha^* 
\mathcal{E}$, where $\alpha: \Sigma \times \mathcal{M} \rightarrow X$
is a universal instanton, and $\pi: \Sigma \times \mathcal{M} \rightarrow
\mathcal{M}$ the projection.  Physically, $\mathcal{F}$ is the sheaf
of $\lambda^a$ zero modes, over the space of bosonic zero modes.

On the (2,2) locus, where $\mathcal{E} = TX$, we have $\mathcal{F} = T
\mathcal{M}$.  (Experts will recall that we are assuming a {\it fixed}
complex structure on the worldsheet.)

When there are no excess ({\it i.e.} $\lambda^a$
or $\psi_+^{\overline{\imath}}$) zero modes, meaning that
\begin{displaymath}
R^1 \pi_* \alpha^* \mathcal{E} \: = \: 0 \: = \:
R^1 \pi_* \alpha^* TX
\end{displaymath}
then Grothendieck-Riemann-Roch tells us that the constraints
\begin{eqnarray*}
\Lambda^{top} \mathcal{E}^{\vee} & = & K_X \\
\mbox{ch}_2(\mathcal{E}) & = & \mbox{ch}_2(TX)
\end{eqnarray*}
imply\footnote{Technically one derives a statement about
first Chern classes from Grothendieck-Riemann-Roch,
which is slightly weaker than the stated implication.
We will ignore this subtlety in this talk.} that
\begin{displaymath}
\Lambda^{top} \mathcal{F}^{\vee} \: \cong \: K_{ \mathcal{M} }
\end{displaymath}
which will be crucial for computing correlation functions.

\section{Quantum correlation functions -- no excess zero modes}

First, let us review the A model calculation.
We will assume there are no $\psi_z^{\overline{\imath}}$ or 
$\psi_{\overline{z}}^i$ zero modes.
Each operator, corresponding to an element of
$H^{p_i,q_i}(X)$, defines an element of $H^{p_i, q_i}(\mathcal{M})$.
Correlation functions can be expressed as
\begin{displaymath}
< \mathcal{O}_1 \cdots \mathcal{O}_m > \: \sim \:
\int_{ \mathcal{M} } 
H^{p_1,q_1}(\mathcal{M}) \wedge \cdots \wedge H^{p_m,q_m}(\mathcal{M}).
\end{displaymath}
The left- and right-moving $U(1)$ anomalies give the selection rules
\begin{displaymath}
\sum p_i \: = \: \sum q_i \: = \: \mbox{dim } \mathcal{M}.
\end{displaymath}
Thus, we can write a correlation function as
\begin{displaymath}
< \mathcal{O}_1 \cdots \mathcal{O}_m > \: \sim \:
\int_{ \mathcal{M} } 
\left( \mbox{top form} \right).
\end{displaymath}

Tree-level correlation function computations in the (0,2) model are similar.
We will assume there are no $\psi_+^{\overline{\imath}}$ or 
$\lambda^a$
zero modes.
Each operator, corresponding to an element of $H^{q_i}\left( X, \Lambda^{p_i}
\mathcal{E}^{\vee} \right)$, defines\footnote{The precise map from
sheaf cohomology on $X$ to sheaf cohomology on $\mathcal{M}$ is discussed
in \cite{ks}, where it is also checked that when $\mathcal{E} = TX$,
this map reduces to the usual map on the (2,2) locus.} 
an element of $H^{q_i}\left( \mathcal{M},
\Lambda^{p_i} \mathcal{F}^{\vee} \right)$.
Omitting a ratio of operator determinants, which at tree level will
only contribute a number, correlation functions can be expressed as
\begin{displaymath}
< \mathcal{O}_1 \cdots \mathcal{O}_m > \: \sim \:
\int_{ \mathcal{M} } 
H^{q_1}\left( \mathcal{M}, \Lambda^{p_1} \mathcal{F}^{\vee} \right) \wedge \cdots
\wedge H^{q_m} \left( \mathcal{M}, \Lambda^{p_m} \mathcal{F}^{\vee} \right).
\end{displaymath}
The left- and right-moving $U(1)$ anomalies give the selection rules
\begin{displaymath}
\sum q_i \: = \: \mbox{dim } \mathcal{M}, \: \: \:
\sum p_i \: = \: \mbox{rank } \mathcal{F}.
\end{displaymath}
Thus, we can write a correlation function as
\begin{displaymath}
< \mathcal{O}_1 \cdots \mathcal{O}_m > \: \sim \:
\int_{ \mathcal{M} } 
H^{top} \left( \mathcal{M}, \Lambda^{top} \mathcal{F}^{\vee} \right).
\end{displaymath}
We saw previously that anomaly cancellation and
Grothendieck-Riemann-Roch imply that $\Lambda^{top} \mathcal{F}^{\vee} \cong
K_{ \mathcal{M} }$, so the integrand above is a top-form, as needed.

\section{Quantum correlation functions -- excess zero modes}

Let us review how to calculate A model correlation functions
when there are $\psi_z^{\overline{\imath}}$ or 
$\psi_{ \overline{z}}^i$
zero modes.  The trick here is to use the four-fermi term
\begin{displaymath}
\int_{ \Sigma} R_{i \overline{\jmath} k \overline{l} } \chi^i 
\chi^{\overline{\jmath}} \psi^k \psi^{\overline{l}}
\end{displaymath}
appearing in the action to soak up the excess zero modes.
For each complex pair of $\psi$ zero modes, we bring down one copy
of the four-fermi term above.
As described by \cite{pauldave,wittenwzw},
the resulting correlation functions have the form
\begin{displaymath}
< \mathcal{O}_1 \cdots \mathcal{O}_m > \: \sim \:
\int_{ \mathcal{M}} H^{\sum p_i, \sum q_i } (\mathcal{M})
\wedge c_{top}(\mbox{Obs})
\end{displaymath}
where $\mbox{Obs}$, known as the ``obstruction sheaf,''
 is the sheaf over $\mathcal{M}$ defined by the
$\psi_z^{\overline{\imath}}$ zero modes.
Mathematically, $\mbox{Obs} = R^1 \pi_* \alpha^* TX$.
The selection rules for left- and right-moving $U(1)$'s say that
\begin{displaymath}
\sum p_i \: + \: \mbox{rank } \mbox{Obs} \: = \:
\sum q_i \: + \: \mbox{rank } \mbox{Obs} \: = \:
\mbox{dim } \mathcal{M}.
\end{displaymath}
Thus, as usual, the integrand in the expression for the correlation
function is a top-form.

Next, let us consider the (0,2) model.
We shall assume that 
\begin{displaymath}
\mbox{rank } R^1 \pi_* \alpha^* \mathcal{E} \: = \:
\mbox{rank } R^1 \pi_* \alpha^* TX \: = \: n.
\end{displaymath}
As in the A model, we shall use the four-fermi term, which now has
the form
\begin{displaymath}
\int_{ \Sigma} F_{i \overline{\jmath} a \overline{b} }
\psi_+^i \psi_+^{\overline{\jmath}} \lambda^a \lambda^{\overline{b}}.
\end{displaymath}
The fermion zero modes define sheaves over the moduli space 
$\mathcal{M}$ as follows:
\begin{displaymath}
\begin{array}{ccccccc}
\psi_+^i & \sim & T \mathcal{M} \: = \: R^0 \pi_*
\alpha^* TX, & \: \: \: &
\lambda^{\overline{a}} & \sim & \mathcal{F} \: = \: R^0 \pi_* \alpha^* \mathcal{E}, \\
\psi_+^{\overline{\imath}} & \sim & \mbox{Obs} \: = \: R^1 \pi_* \alpha^* TX, & \: \: \: &
\lambda^a & \sim & \mathcal{F}_1 \: \equiv \:
R^1 \pi_* \alpha^* \mathcal{E}.
\end{array}
\end{displaymath}
If we think about the four-fermi term as a bundle-valued differential
form on moduli space, then on symmetry grounds we are led to identify
each four-fermi term with an element of
\begin{displaymath}
H^1\left( \mathcal{M}, \mathcal{F}^{\vee} \otimes \mathcal{F}_1 \otimes
\left( \mbox{Obs} \right)^{\vee} \right).
\end{displaymath}

Thus, bringing down $n$ four-fermi terms to absorb all excess zero modes,
correlation functions can be written
\begin{displaymath}
< \mathcal{O}_1 \cdots \mathcal{O}_m > \: \sim \:
\int_{ \mathcal{M} } H^{\sum q_i }\left( \mathcal{M},
\Lambda^{\sum p_i} \mathcal{F}^{\vee} \right) \wedge
H^n\left( \mathcal{M}, \Lambda^n \mathcal{F}^{\vee} \otimes
\Lambda^n \mathcal{F}_1 \otimes \Lambda^n \left( \mbox{Obs} \right)^{\vee}
\right)
\end{displaymath}
where $n = \mbox{rank } \mathcal{F}_1 = \mbox{rank } \mbox{Obs}$.
The left- and right-$U(1)$ anomalies give the selection rules
\begin{displaymath}
\sum q_i \: + \: n \: = \: \mbox{dim } \mathcal{M}, \: \: \:
\sum p_i \: + \: n \: = \: \mbox{rank } \mathcal{F}.
\end{displaymath}

The first thing to check is that the integrand above is a top-form.
Previously, when there were no excess zero modes, we saw that
anomaly cancellation and Grothendieck-Riemann-Roch imply that
$\Lambda^{top} \mathcal{F}^{\vee} \cong K_{ \mathcal{M} }$.
When there are excess zero modes, the result is modified.
Given the two assumptions on $\mathcal{E}$, namely
\begin{eqnarray*}
\Lambda^{top} \mathcal{E}^{\vee} & \cong & K_X, \\
\mbox{ch}_2 (\mathcal{E}) & = & \mbox{ch}_2(TX),
\end{eqnarray*}
Grothendieck-Riemann-Roch now implies that
\begin{displaymath}
\Lambda^{top} \mathcal{F} \otimes \Lambda^{top} \mathcal{F}_1^{\vee} \: \cong \:
\Lambda^{top} T \mathcal{M} \otimes \Lambda^{top} \left(
\mbox{Obs} \right)^{\vee}
\end{displaymath}
or, phrased more simply,
\begin{displaymath}
\Lambda^{top} \mathcal{F}^{\vee} \otimes \Lambda^{top} \mathcal{F}_1
\otimes \Lambda^{top} \left( \mbox{Obs} \right)^{\vee} \: \cong \:
K_{ \mathcal{M} }
\end{displaymath}
which tells us that the integrand of our expression for the
correlation function is, again, a top-form.

Having established that our interpretation of the four-fermi term
does indeed lead to correlation functions expressable as
integrals of top-forms, we now need to check that our
(0,2) expressions reduce to the A model expression when
$\mathcal{E} = TX$.  In particular, we need a relation between
Chern classes and sheaf cohomology.

Such a relationship is provided by Atiyah classes \cite{atiyah}.
Consider the curvature $F$ of a connection on a holomorphic bundle
$\mathcal{E}$ on $X$.  $F$ is an endomorphism-valued $(1,1)$-form.
Since it is a holomorphic connection, {\it i.e.}
it has no $(2,0)$ or $(0,2)$ parts,
the Bianchi identity $d F = 0$ implies that $F$ is 
$\overline{\partial}$-closed, and so represents an element of the
sheaf cohomology group
\begin{displaymath}
H^1\left(X, \Omega^1_X \otimes \mathcal{E}^{\vee} \otimes \mathcal{E} \right).
\end{displaymath}
The element of this sheaf cohomology group corresponding to $F$
is known as the Atiyah class of $\mathcal{E}$.
Since Chern characters of $\mathcal{E}$ can be expressed as
traces over products of $F$ with itself, we see that Chern classes
are encoded in products of the sheaf cohomology group above with itself.

Returning to our (0,2) model discussion,
when $\mathcal{E} = TX$, each four-fermi term generates a factor lying in
\begin{displaymath}
H^1\left( \mathcal{M}, \mathcal{F}^{\vee} \otimes \mathcal{F}_1
\otimes \left( \mbox{Obs} \right)^{\vee} \right) \: = \:
H^1\left( \mathcal{M}, \Omega^1_{ \mathcal{M} } \otimes 
\left( \mbox{Obs} \right)^{\vee} \otimes \mbox{Obs} \right)
\end{displaymath}
which is the same sheaf cohomology group that contains the
Atiyah class of the obstruction sheaf $\mbox{Obs}$.

Furthermore, bringing down $n = \mbox{rank }\mbox{Obs}$ factors of
the sheaf cohomology group above generates a factor of
\begin{displaymath}
H^n\left( \mathcal{M}, \Omega^n_{ \mathcal{M} } \otimes
\Lambda^{top} \left( \mbox{Obs} \right)^{\vee} \otimes
\Lambda^{top} \mbox{Obs} \right)
\end{displaymath}
which contains $c_{top}( \mbox{Obs} )$.

Thus, our proposed interpretation of the (0,2) four-fermi term
not only generates top forms, but also correctly reproduces
the (2,2) obstruction bundle story.

\section{Compactifications of $\mathcal{M}$}

In order to make sense of expressions such as
\begin{displaymath}
\int_{ \mathcal{M} } \left( \mbox{top form} \right)
\end{displaymath}
we need the space of bosonic zero modes $\mathcal{M}$ to be compact.

However, it is well-known that spaces of honest holomorphic maps
are {\it not} compact.  For example, the space of honest degree 1 maps
${\bf P}^1 \rightarrow {\bf P}^1$ is the group manifold of $SL(2,{\bf C})$,
which is certainly not compact.

In order to make sense of our expressions for correlation functions,
we must compactify $\mathcal{M}$.
Furthermore, in the (0,2) case, we also need to extend the sheaves
$\mathcal{F}$, $\mathcal{F}_1$ over the compactification,
in a way that preserves the crucial property
\begin{displaymath}
\Lambda^{top} \mathcal{F}^{\vee} \otimes
\Lambda^{top} \mathcal{F}_1 \otimes
\Lambda^{top} \left( \mbox{Obs} \right)^{\vee}
\: \cong \:
K_{ \mathcal{M} }
\end{displaymath}
(generalizing $\Lambda^{top} \mathcal{F}^{\vee} \cong K_{ \mathcal{M} }$
to the case of excess zero modes)
that we derived from the anomaly cancellation condition,
using Grothendieck-Riemann-Roch.

It turns out that gauged linear sigma models not only naturally
compactify $\mathcal{M}$, as discussed in \cite{daveronen1},
but also naturally extend $\mathcal{F}$, $\mathcal{F}_1$ in the desired form
\cite{ks}, as we shall review here.

\section{Review of gauged linear sigma models}

A (2,2) gauged linear sigma model describes toric varieties
as chiral superfields
with gauged $U(1)$ actions.
Each (2,2) chiral superfield contains a complex boson $\phi$,
left- and right-moving complex fermions $\psi_-$, $\psi_+$,
and an auxiliary field $F$.
For example, we can describe ${\bf P}^{N-1}$ as
$N$ chiral superfields, each of charge one with respect to a gauged
$U(1)$.
The D-term condition from the gauged $U(1)$ gives the constraint
\begin{displaymath}
\sum | \phi_i |^2 \: = \: r
\end{displaymath}
which restricts the Higgs moduli space to a sphere $S^{2N-1}$.
Modding out the gauge symmetry leaves us with
$S^{2N-1}/U(1) = {\bf P}^{N-1}$.

Calabi-Yau manifolds obtained as complete intersections in toric
varieties can be described by adding superpotential terms;
the zero locus of the bosonic potential is the Calabi-Yau.
In this talk, however, we will only be concerned with massive
theories describing toric varieties (and bundles thereon).

A (0,2) gauged linear sigma model \cite{drev,dk}
is constructed primarily with
two types of superfields:
\begin{itemize}
\item The (0,2) chiral superfield contains a complex boson $\phi$
and a right-moving complex fermion $\psi_+$.
\item The (0,2) fermi superfield contains a left-moving complex fermion
$\psi_-$ and an auxiliary field $F$.
\end{itemize}
Together, these two (0,2) superfields form a (2,2) chiral superfield.

The fermi superfields have an important quirk:
although $\overline{D}_+ \Phi = 0$ for $\Phi$ chiral,
one can have $\overline{D}_+ \Lambda = E$ for $\Lambda$ fermi,
where $E$ is a chiral superfield ($\overline{D}_+ E = 0$).
This constrains the superpotential, and plays an important role
in describing some kinds of bundles, as we shall review momentarily.

A toric variety can be described in (0,2) language as a collection
of (0,2) chiral superfields with some gauged $U(1)$'s, just as in
the (2,2) case.

There are several ways to describe bundles over the toric variety.
The two types we shall discuss in this short note are:
\begin{itemize}
\item Reducible bundles.  Although rarely seen in practice,
the easiest kind of bundle to describe is a completely reducible
bundle, of the form
\begin{displaymath}
\mathcal{E} \: = \: \oplus_a \mathcal{O}(\vec{n}_a)
\end{displaymath}
In a (0,2) gauged linear sigma model this is described by
a collection of fermi superfields $\Lambda^a$,
with charges $\vec{n}_a$ under the gauged $U(1)$'s, that are otherwise
free.
\item Cokernel.  Another common realization of bundles $\mathcal{E}$
is as cokernels:
\begin{displaymath}
0 \: \longrightarrow \: \mathcal{O}^{\oplus k} \:
\stackrel{ E_a^{\lambda} }{\longrightarrow} \:
\oplus_a \mathcal{O}(\vec{n}_a) \: \longrightarrow \:
\mathcal{E} \: \longrightarrow \: 0
\end{displaymath}
Here, we have fermi superfields $\Lambda^a$ with charges $\vec{n}_a$,
plus $k$ neutral chiral superfields $\Sigma^{\lambda}$,
with $\overline{D}_+ \Lambda^a = \Sigma^{\lambda} E^a_{\lambda}(\phi)$.
\end{itemize}
It is also possible to physically build bundles as kernels and,
more generally, as the cohomology of a monad, but we leave discussion
of those presentations to \cite{ks}.

Anomaly cancellation in a (0,2) gauged linear sigma model is slightly
stronger than the usual statement in a large-radius nonlinear sigma model.
If we let $\vec{n}_a$ denote charges of left-moving fermions and
$\vec{q}_i$ denote charges of right-moving fermions, then the anomaly
cancellation condition in the (0,2) gauged linear sigma model is
\begin{displaymath}
\sum_a n_a^t n_a^s \: = \: \sum_i q_i^t q_i^s
\end{displaymath}
for each $s$, $t$, which implies, but is slightly stronger than,
$\mbox{ch}_2(\mathcal{E}) = \mbox{ch}_2(TX)$.
(In fact, this linear sigma model anomaly cancellation condition
is strong enough to distinguish different presentations of the same
bundle, as discussed in \cite{ks}.)
We shall also assume that
\begin{displaymath}
\sum_a n_a^t \: = \: \sum q_i^t
\end{displaymath}
for each $t$, which implies, but is slightly stronger than,
$c_1(\mathcal{E}) = c_1(TX)$.

\section{Linear sigma model compactifications}

The basic idea of \cite{daveronen1} in constructing linear-sigma-model-based
compactifications of moduli spaces is to expand the fields of the linear
sigma model in a basis of zero modes, then the coefficients in that expansion
form homogeneous coordinates on the compactified moduli space.

For example, consider a gauged linear sigma model describing
${\bf P}^{N-1}$, {\it i.e.}, $N$ chiral superfields $x_i$,
each of charge one with respect to a gauged $U(1)$.
The gauge instantons of the gauged linear sigma model become the worldsheet
instantons of the corresponding nonlinear sigma model.
In this example, to describe the moduli space of degree $d$ maps
${\bf P}^1 \rightarrow {\bf P}^{N-1}$, we expand 
\begin{eqnarray*}
x_i & \in & \Gamma\left( \mathcal{O}(1 \cdot d) \right) \\
& = & x_{i0} u^d \: + \: x_{i1} u^{d-1} v \: + \: \cdots \: + \:
x_{id} v^d
\end{eqnarray*}
where $u$, $v$ are homogeneous coordinates on the worldsheet
(${\bf P}^1$).  The $(x_{ij})$ are homogeneous coordinates on the
moduli space $\mathcal{M}$.  We omit the point where all the
$x_i \equiv 0$, and give each coordinate $x_{ij}$ the same $U(1)$ charge
as $x_i$.  As a result,
\begin{displaymath}
\mathcal{M} \: = \: {\bf P}^{N(d+)-1}.
\end{displaymath}

More generally, given chiral superfields $x_i$ of charges $\vec{q}_i$,
a moduli space of maps of degree $\vec{d}$ is determined by taking
\begin{displaymath}
x_i \: \in \: \Gamma\left( \mathcal{O}(\vec{q}_i \cdot \vec{d} ) \right)
\end{displaymath}
and following the same general procedure as for ${\bf P}^{N-1}$ above.

The same ideas allow us to induce bundles on linear sigma model
moduli spaces.

Just as worldsheet fields define line bundles on the target toric
variety, we can expand those worldsheet fields in a basis of zero modes,
and take the coefficients to define line bundles on $\mathcal{M}$.

For example, suppose we have a completely reducible bundle
\begin{displaymath}
\mathcal{E} \: = \: \oplus_a \mathcal{O}(\vec{n}_a)
\end{displaymath}
which physically is described by a collection of fermi superfields
$\Lambda^a$ with charges $\vec{n}_a$.  We expand each left-moving fermion
$\lambda_-^a$ in a basis of zero modes, as
\begin{eqnarray*}
\lambda_-^a & \in & \Gamma\left( \mathcal{O}( \vec{n}_a \cdot \vec{d} ) \right) \\
& = & \lambda_-^{a0} u^{\vec{n}_a \cdot \vec{d} + 1} \: + \:
\lambda_a^{a1} u^{\vec{n}_a \cdot \vec{d}} v \: + \: \cdots
\end{eqnarray*}
(assuming that $\vec{n}_a \cdot \vec{d} \geq 0$),
and take each $\lambda_-^{ai}$ to correspond to a line bundle
$\mathcal{O}(\vec{n}_a)$ on $\mathcal{M}$.

Thus, we get the induced bundle
\begin{displaymath}
\mathcal{F} \: = \: \oplus_a H^0\left( {\bf P}^1, \mathcal{O}(\vec{n}_a \cdot
\vec{d} ) \right) \otimes_{ {\bf C} } \mathcal{O}(\vec{n}_a). 
\end{displaymath}
Similarly, we can also derive
\begin{displaymath}
\mathcal{F}_1 \: = \: \oplus_a H^1\left( {\bf P}^1, \mathcal{O}(\vec{n}_a \cdot
\vec{d} ) \right) \otimes_{ {\bf C} } \mathcal{O}(\vec{n}_a).
\end{displaymath}

We need to check that this ansatz has the correct ranks and $c_1$'s,
but before we perform those checks for the reducible bundle case,
let us first go on to the cokernel case, and then return to the
reducible bundle case afterwards.

Suppose $\mathcal{E}$ is described as a cokernel:
\begin{displaymath}
0 \: \longrightarrow \:
\mathcal{O}^{\oplus m} \: \stackrel{E}{\longrightarrow} \:
\oplus_a \mathcal{O}(\vec{n}_a ) \: \longrightarrow \mathcal{E} \:
\longrightarrow \: 0.
\end{displaymath}
In addition to fermi superfields $\Lambda^a$ with charges as above,
recall we also have $m$ neutral chiral superfields $\Sigma^{\lambda}$.
As before, we expand the left-moving fermions $\lambda_-^a$ and the
bosons $\sigma^{\lambda}$ in a basis of zero modes, and interpret
the coefficients as defining line bundles on $\mathcal{M}$.

Expanding in zero modes as before, and also expanding the maps
in zero modes to get maps between the induced line bundles, we find
\begin{displaymath}
0 \: \longrightarrow \: \oplus_1^m H^0\left( {\bf P}^1, \mathcal{O}(\vec{0} \cdot
\vec{d} ) \right) \otimes_{ {\bf C} } \mathcal{O} \: \longrightarrow \:
\oplus_a H^0\left( {\bf P}^1, \mathcal{O}(\vec{n}_a \cdot \vec{d}) \right)
\otimes_{ {\bf C} } \mathcal{O}(\vec{n}_a ) \: \longrightarrow \:
\mathcal{F}.
\end{displaymath}

Now, in this particular case, it turns out that the last map, into $\mathcal{F}$,
is surjective.  In general, however, it will not be surjective, and in any
event, there is further information to be gained.  
In general, we have a (long) exact sequence given by
\begin{displaymath}
\begin{array}{ccccccc}
0 & \longrightarrow & \oplus_1^m H^0\left( {\bf P}^1, \mathcal{O}(\vec{0} \cdot
\vec{d} ) \right) \otimes_{ {\bf C} } \mathcal{O} & \longrightarrow &
\oplus_a H^0\left( {\bf P}^1, \mathcal{O}(\vec{n}_a \cdot \vec{d}) \right)
\otimes_{ {\bf C} } \mathcal{O}(\vec{n}_a ) & \longrightarrow &
\mathcal{F} \\
 & \longrightarrow &
\oplus_1^m H^1\left( {\bf P}^1, \mathcal{O}(\vec{0} \cdot \vec{d} ) \right)
\otimes_{ {\bf C} } \mathcal{O} & \longrightarrow &
\oplus_a H^1\left( {\bf P}^1, \mathcal{O}(\vec{n}_a \cdot \vec{d}) \right)
\otimes_{ {\bf C} } \mathcal{O}(\vec{n}_a ) & \longrightarrow &
\mathcal{F}_1 \\
 & \longrightarrow & 0 & & & & 
\end{array}
\end{displaymath}
which simplifies in the present case to give
\begin{displaymath}
0 \: \longrightarrow \: \mathcal{O}^{\oplus m} \: \longrightarrow \:
\oplus_a H^0\left( {\bf P}^1, \mathcal{O}(\vec{n}_a \cdot \vec{d} ) \right)
\otimes_{ {\bf C} } \mathcal{O}(\vec{n}_a ) \: \longrightarrow \:
\mathcal{F} \: \longrightarrow \: 0
\end{displaymath}
and
\begin{displaymath}
\mathcal{F}_1 \: \cong \: \oplus_a H^1\left( {\bf P}^1, \mathcal{O}(\vec{n}_a \cdot
\vec{d}) \right) \otimes_{ {\bf C} } \mathcal{O}(\vec{n}_a ).
\end{displaymath}

Let us check that this gives correct results.

First, let us check that when $\mathcal{E} = TX$,
{\it i.e.} the (2,2) locus,
the induced sheaf $\mathcal{F} = T \mathcal{M}$.
To do this, note that if our toric variety is defined by homogeneous
coordinates with charges $\vec{q}_i$ with respect to $k$ $U(1)$'s,
then the tangent bundle can be expressed as
\begin{displaymath}
0 \: \longrightarrow \: \mathcal{O}^{\oplus k} \: \longrightarrow \:
\oplus_q \mathcal{O}(\vec{q}_i) \: \longrightarrow \: TX \:
\longrightarrow \: 0.
\end{displaymath}
The results above are immediately applicable, and we see immediately
that the induced sheaf $\mathcal{F}$ is precisely the tangent bundle
of the linear sigma model moduli space described earlier.
It can also be shown \cite{ks} that when $\mathcal{E} = TX$,
the sheaf $\mathcal{F}_1$ coincides with the obstruction sheaf.

Let us check that more generally, the induced sheaves $\mathcal{F}$,
$\mathcal{F}_1$ have reasonable properties.
It can be shown \cite{ks} that they match $R^0 \pi_* \alpha^* \mathcal{E}$,
$R^1 \pi_* \alpha^* \mathcal{E}$ on the open subset of $\mathcal{M}$
corresponding to honest maps.  Let us check that their first Chern classes
have the desired property.

From the expressions for $\mathcal{F}$, $\mathcal{F}_1$ above,
we have that
\begin{eqnarray*}
c_1(\mathcal{F}) \: - \: c_1\left( \mathcal{F}_1 \right) & = &
\left[ \sum_{\vec{n}_a \cdot \vec{d} \geq 0} \left( \vec{n}_a \cdot \vec{d}
\: + \: 1 \right) n_a^t J_t \right] \: - \:
\left[ \sum_{\vec{n}_a \cdot \vec{d} < 0 } \left( - \vec{n}_a \cdot \vec{d}
\: - \: 1 \right) n_a^t J_t \right]\\
& = & \left[ \sum_a n_a^t J_t \right] \: + \: \left[
\sum_a \left( \vec{n}_a \cdot \vec{d} \right) n_a^t J_t \right]
\end{eqnarray*}
where the $J_t$ generate $H^2({\bf Z})$.
Now, if we express the tangent bundle in the form
\begin{displaymath}
0 \: \longrightarrow \: \mathcal{O}^{\oplus k} \: \longrightarrow \:
\oplus_i \mathcal{O}(\vec{q}_i) \: \longrightarrow \: TX \:
\longrightarrow \: 0
\end{displaymath}
then linear sigma model anomaly cancellation and the linear sigma
model version of the constraint $\Lambda^{top} \mathcal{E}^{\vee} \cong K_X$
give us the constraints
\begin{eqnarray*}
\sum_a n_a^t & = & \sum_i q_i^t \: \mbox{ for all } t, \\
\sum_a n_a^t n_a^s & = & \sum_i q_i^t q_i^s \: \mbox{ for all } s, t.
\end{eqnarray*}
Plugging these constraints into the expression above for first Chern classes,
we find that
\begin{eqnarray*}
c_1(\mathcal{F}) \: - \: c_1\left( \mathcal{F}_1 \right) & = &
\left[ \sum_i q_i^t J_t \right] \: + \:
\left[ \sum_i \left( \vec{q}_i \cdot \vec{d} \right) q_i^t J_t \right]\\
& = &
c_1 \left( T \mathcal{M} \right) \: - \: c_1 \left( \mbox{Obs} \right).
\end{eqnarray*}
Thus\footnote{In making the next statement we are assuming that
$\mathcal{M}$ is smooth, K\"ahler, and simply-connected.},
\begin{displaymath}
\Lambda^{top} \mathcal{F} \otimes \Lambda^{top} \mathcal{F}_1 \: = \:
\Lambda^{top} T \mathcal{M} \otimes \Lambda^{top} \left( \mbox{Obs} \right)^{\vee}
\end{displaymath}
exactly as desired.

Thus, not only does the gauged linear sigma model naturally define extensions of
$R^{0,1}\pi_* \alpha^* \mathcal{E}$ across the compactification divisor,
but those extensions have good properties.

Let us now return to the case of reducible bundles, and perform the same test.
Recall that if $\mathcal{E} = \oplus_a \mathcal{O}(\vec{n}_a)$, then
\begin{eqnarray*}
\mathcal{F} & \cong & \oplus_a H^0\left( {\bf P}^1, \mathcal{O}(\vec{n}_a \cdot \vec{d})
\right) \otimes_{ {\bf C} } \mathcal{O}(\vec{n}_a), \\
\mathcal{F}_1 & \cong & \oplus_a H^1\left( {\bf P}^1,
\mathcal{O}(\vec{n}_a \cdot \vec{d} ) \right) \otimes_{ {\bf C} }
\mathcal{O}(\vec{n}_a ).
\end{eqnarray*}
It can be shown \cite{ks} that these sheaves agree with
$R^{0,1} \pi_* \alpha^* \mathcal{E}$ on the open subset of $\mathcal{M}$
corresponding to honest maps.
Let us check first Chern classes.
\begin{eqnarray*}
c_1(\mathcal{F}) \: - \: c_1\left( \mathcal{F}_1 \right) & = &
\left[ \sum_{ \vec{n}_a \cdot \vec{d} \geq 0 } \left( \vec{n}_a \cdot \vec{d}
\: + \: 1 \right) n_a^t J_t \right] \: - \:
\left[ \sum_{ \vec{n}_a \cdot \vec{d} < 0 } \left( - \vec{n}_a \cdot \vec{d}
\: - \: 1 \right) n_a^t J_t \right] \\
& = & \left[ \sum_a n_a^t J_t \right] \: + \:
\left[ \sum_a \left( \vec{n}_a \cdot \vec{d} \right) n_a^t J_t \right].
\end{eqnarray*}
Using the same conventions for the tangent bundle as previously,
the linear sigma model consistency conditions imply that
\begin{eqnarray*}
c_1(\mathcal{F}) \: - \: c_1 \left( \mathcal{F}_1 \right) & = &
\left[ \sum_i q_i^t J_t \right] \: + \:
\left[ \sum_i \left( \vec{q}_i \cdot \vec{d} \right) q_i^t J_t \right] \\
& = & c_1 ( T \mathcal{M} ) \: - \: c_1\left( \mbox{Obs} \right)
\end{eqnarray*}
exactly as desired.

Our description of $\mathcal{F}$, $\mathcal{F}_1$ is presentation-dependent,
and as the reader may have guessed, different physical presentations
of the same bundle $\mathcal{E}$ can lead to different extensions of
$R^{0,1} \pi_* \alpha^* \mathcal{E}$ over the compactification divisor,
{\it i.e.} different $\mathcal{F}$, $\mathcal{F}_1$.
This is discussed in detail in \cite{ks}.

\section{Conclusions}

In this short note we have outlined some of the results of
\cite{ks}, generalizing the rational curve counting of the A model
to perturbative heterotic strings.  After formally outlining how
the relevant correlation functions can be defined mathematically,
we outlined how linear sigma models not only naturally compactify
moduli spaces, but also extend needed sheaves over those compactifications.
Further details, and a specific computation verifying results of
\cite{adamsbasusethi}, can be found in \cite{ks}.

\bibliographystyle{amsalpha}

\end{document}